\documentclass[aps,prb,twocolumn,superscriptaddress,showpacs]{revtex4-1}

\usepackage{graphicx,psfrag}
\usepackage[latin1]{inputenc}
\usepackage{natbib}
\usepackage{amsmath, amsthm, amssymb}
\usepackage{color}
\usepackage{soul}
\usepackage{comment}
\usepackage{hyperref}
\usepackage{lineno}
\usepackage{amsfonts}
\usepackage{fancyhdr}
\usepackage{bm}

\newcommand{\mb}{\mathbf}
\newcommand{\notes}[1]{}

\newcommand{\beq}{\begin{equation}}
\newcommand{\eeq}{\end{equation}}
\newcommand{\beqnn}{\begin{equation*}}
\newcommand{\eeqnn}{\end{equation*}}
\newcommand{\beqas}{\begin{eqnarray*}}
	\newcommand{\eeqas}{\end{eqnarray*}}
\newcommand{\beqa}{\begin{eqnarray}}
\newcommand{\eeqa}{\end{eqnarray}}
\newcommand{\p}{\partial}

\begin{document}

\title{Effects of the Zhang-Li Torque on Spin Torque nano Oscillators}

\author{Jan Albert}
\affiliation{Dept.\ of Condensed Matter Physics, University of Barcelona, 08028 Barcelona, Spain}
\affiliation{Department of Physics and Astronomy, Stony Brook University, Stony Brook, New York 11794, USA}
\author{Ferran Maci\`a}
\email{ferran.macia@ub.edu}
\affiliation{Dept.\ of Condensed Matter Physics, University of Barcelona, 08028 Barcelona, Spain}
\affiliation{Institute of Nanoscience and Nanotechnology (IN2UB), University of Barcelona, 08028 Barcelona, Spain}

\author{Joan Manel Hern\`andez}
\affiliation{Dept.\ of Condensed Matter Physics, University of Barcelona, 08028 Barcelona, Spain}
\affiliation{Institute of Nanoscience and Nanotechnology (IN2UB), University of Barcelona, 08028 Barcelona, Spain}

\date{\today}

\begin{abstract}
	Spin-torque nano-oscillators (STNO) are microwave auto-oscillators based on magnetic resonances having a nonlinear response with the oscillating amplitude, which provides them with a large frequency tunability including the possibility of mutual synchronization. The magnetization dynamics in STNO are induced by spin transfer torque (STT) from spin currents and can be detected by changes in electrical resistance due to giant magnetoresistance or tunneling magnetoresistance. The STT effect is usually treated as a damping-like term that reduces magnetic dissipation and promotes excitation of magnetic modes. However, an additional term, known as Zhang-Li term has an effect on magnetization gradients such as domain walls, and could have an effect on localized magnetic modes in STNO. Here we study the effect of Zhang-Li torques in magnetic excitations produced in STNO with a nanocontact geometry. Using micromagnetic simulations we find that Zhang-Li torque modify threshold currents of magnetic modes and their effective sizes. Additionally we show that effects can be controlled by changing the ratio between nanocontact size and layer thickness.
\end{abstract}

\maketitle



\section{Introduction} \label{introduction}
Magnetic nano-oscillators are a popular field of study given their potential in a wide range of applications; from conventional electronics using microwave signal elements to neuromorphic computing schemes. Auto oscillations at gigahertz frequencies can be achieved in magnetic systems by the spin-transfer torque (STT) effect \cite{Tsoi1998}, where a flow of spin angular momentum can compensate dissipation resulting in magnetic excitations. Spin-torque nano-oscillators (STNO) consisting of a point contact to a thin film ferromagnet (FM), were first proposed theoretically \cite{Slonczewski1999} where a dc current density generated a high-frequency dynamic response in a FM layer resulting in spin-wave emission. Subsequent devices with different geometries of magnetic layers have emerged and showed the possibility of generating a wide variety of magnetic excitations---including spin wave radiation \cite{Kiselev2003,Rippard2004} and localized magnetic modes such as vortices \cite{Vortex_heyne,Vortex_nature}, bullets \cite{SlavinBullet,Madami2011}, and dissipative droplet solitons \cite{Mohseni2013,Macia2014} among others \cite{Demidov2010,Bonetti2015}.

Transfer of a spin angular momentum flow into a magnetic system can occur from a spin-polarized charge current \cite{Slonczewski1996,Berger1996}, or from a pure spin current \cite{Kato1910_SHE,Wunderlich_SHE}. In the first case, charge currents are polarized using an additional magnetic layer---a polarizer layer (PL)---whereas in the second case charge currents are converted to pure spin currents through the spin Hall effect in a nearby non-magnetic layer. In both cases there are relatively large charge current densities involved---often localized in very small areas \cite{Dumas_STO}.

The reduced magnetization $\mb{m}$ of the free magnetic layer (FL) obeys the Landau-Lifshitz dynamical equation, \begin{equation}
\dot{\mb{m}}(\mb{r},t) = \bm{\tau}_{\textrm{LL}} + \bm{\tau}_{\textrm{SL}} + \bm{\tau}_{\textrm{ZL}},
\label{eq:partialm}
\end{equation} 
whose contributions can be broken down into three torques: the Landau-Lifshitz (LL) one (which contains the usual precession and damping terms induced by the external field) and two spin-transfer torques; the Slonczewski (SL) \cite{Slonczewski1996} and Zhang-Li (ZL) \cite{ZhangLi_prb,ZhangLi_prl}.

The SL torque is the main responsible for the generation of magnetic excitations since it can oppose the LL damping term and eventually reverse magnetic moments when the intensity of spin-polarized current is high enough. On the other hand, the ZL torque is related to magnetic gradients and can be seen as a modulation of the generated excitations. 

Nanopillar geometries (where the ferromagnetic layers are also confined) ensure a uniform electric current throughout the multilayer magnetic stack. However, in nanocontact devices, the current injection is confined to a nanoscopic region and it immediately diffuses when entering into the magnetic layer---which extends further than the nanocontact region. Experimental studies of STNO using nanocontacts have shown a variety of both localized and propagating excitations \cite{Brataas2012,Dumas_STO} and micromagnetic simulations \cite{Consolo_prb2007} have been used to describe the observed results and link them with existent theories \cite{Slavin_ieee, Hoefer2010}. However, most of them neglected the effects of the ZL torque. In fact, even when using a micromagnetic simulation software that accounts for the ZL torques, assuming a uniform charge current distribution---as is the case in nanopillar geometries---results in a cancellation of the ZL effects. Chung \textit{et al.} \cite{AkermanXMCD} showed images of magnetic droplet solitons in STNO and suggested that ZL torques might have been the origin of the observed larger-than-expected sizes. Here, we have accurately simulated the dynamics of magnetization in STNO with a nanocontact geometry to study the effects of ZL torques on different types of magnetic excitations.

\section{Dynamical magnetic modes} \label{introduction}

\begin{figure}[h!]
\includegraphics[width=1\columnwidth]{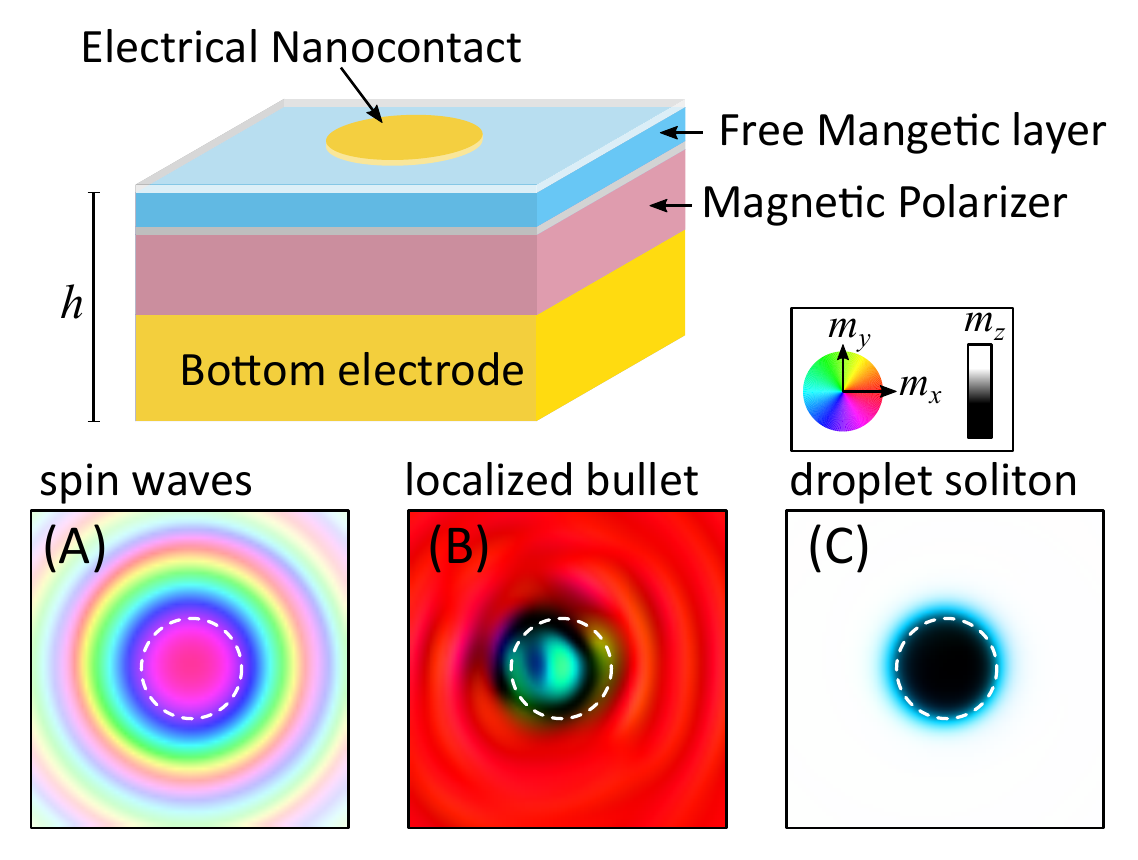}
\caption{\small{Schematic representation of a STNO and associated dynamical magnetic modes. In (A) propagating spin waves, in (B) a localized bullet and in (C)  droplet soliton}}
\label{fig:stno}
\end{figure}

Depending on the orientation of the FL magnetization and the values of mangetic anisotropy, different types of fundamental spin-wave excitations can be generated in STNO with a nanocontact geometry \cite{Dumas_STO} (See Fig.\ \ref{fig:stno}).

When the magnetic layer has easy-plane anisotropy (i.e., the equilibrium magnetization lies in the film plane with no preferred direction) two main spin-wave excitation occur depending on the direction of the magnetization: 

{\it A) Propagating Spin waves} are generated when the magnetic layer is magnetized perpendicularly to the film plane \cite{Slonczewski1999}. The resulting spin waves have a wavelength proportional to the nanocontact radius \cite{Madami2011}.

{\it B) Localized bullets} are generated when the layer's magnetization is in the film's plane as well as the applied magnetic field. In this case, the excited spin-wave mode is strongly nonlinear and self-localized \cite{Slavin_ieee}. The spatial extension of the mode is related to the nanocontact size\cite{Consolo_prb2007}. The effect of Oersted fields or fringe fields might promote other localized modes \cite{Bonetti2015} including vortices.

When the magnetic layer shows uniaxial magnetic anisotropy---in particular in films with perpendicular magnetic anisotropy (PMA)---another fundamental magneto-dynamical mode has been predicted an observed:

{\it C) Magnetic droplet solitons} are nonlinear localized wave excitations consisting of partially reversed precessing magnetization \cite{Hoefer2010,Mohseni2013,Macia2014}. The size of droplets is set by the nanocontact region where damping is suppressed through the STT effect. Droplets have been experimentally created using the STT effect in electric nanocontacts to PMA films \cite{Mohseni2013,Macia2014,Backes2015,AkermanXMCD}. Here the Oersted fields also provide one more degree of complexity and could promote the existence of topological droplets \cite{Zhou_natComm_2015, Sergei2015, Statuto_2018}.

\section{The Zhang-Li Torque}
In a first order approximation, the torque introduced by Zhang and Li in 2004 \cite{ZhangLi_prl} can be considered adiabatic, and then it can be written as
\begin{align}
\bm{\tau}_{\textrm{ZL}} = -\gamma\frac{\hbar}{2eM_{\textrm{sat}}}\frac{1}{1+\alpha^2}\bigg[ \mb{m}\times &\big(\mb{m}\times(\mb{j} \,\cdot \bm{\nabla})\mb{m}\big) \nonumber \\
- \alpha \, &\big(\mb{m}\times(\mb{j} \,\cdot \bm{\nabla})\mb{m}\big) \bigg],
\label{eq:ZL}
\end{align}
where $\gamma$ is the electron gyromagnetic ratio, $M_{\textrm{sat}}$ is the magnetic layer saturation magnetization, $\alpha$ is the damping parameter and $\mb{j}$ is the electric current density. Remaining factors are universal constants.

Given the cylindrical symmetry of point contact STNO, it is convenient to work in the corresponding coordinate system $(\rho,\theta,z)$. Besides, this symmetry leads to $j_{\theta}=0$ (as will be discussed subsequently) and, additionally, the magnetic layer can be considered thin enough to neglect variations of the magnetization along $z$. With these assumptions we can simplify the cross product of Eq.\ \ref{eq:ZL} and obtain
\begin{equation}
|\bm{\tau}_{\textrm{ZL}_1}| \propto j_{\rho}\left|\frac{\p \mb{m}}{\p \rho}\right|,
\label{eq:simplifZL}
\end{equation}
being $|\bm{\tau}_{\textrm{ZL}_1}|$ the modulus of the first term (i.e., the most significant one) of Eq.\ \ref{eq:ZL}. We notice that the effect of ZL torque in the studied case is stronger when the magnetization variation is along the direction of the radial current, $j_{\rho}$, (e.g., a domain wall experiencing an in-plane current).

\begin{figure}[htb!]
	\includegraphics[width=1\columnwidth]{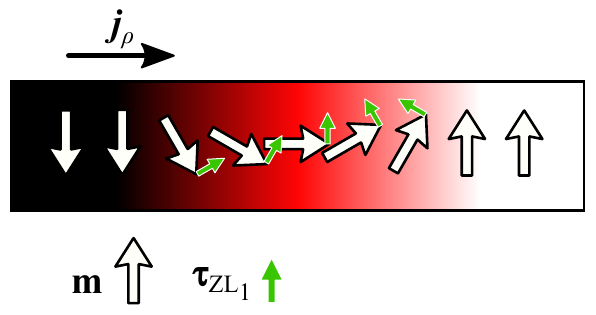}
	\caption{\small{Directions of the first term of Eq.\ \ref{eq:ZL} ($\bm{\tau}_{\text{ZL}_1}$) along a chain of reversing magnetic moments ($\mb{m}$), considering $j_\rho > 0$.}}
	\label{fig:ZLtorque}
\end{figure}

Figure \ref{fig:ZLtorque} shows the direction of $\bm{\tau}_{\textrm{ZL}_1}$ along a given magnetic gradient, which could represent a section of a droplet domain wall. Notice how, when $j_\rho > 0$ (imagine the origin of coordinates is at the left hand side of the schematic plot), the ZL torque favors upwards magnetic moments. If $j_\rho$ changed its sign, $\bm{\tau}_{\textrm{ZL}_1}$ would flip, favoring downward magnetic moments. It is thus reasonable to expect the ZL torque to have an impact on magnetic domain walls; with opposite effects for opposite current directions.

When it comes to simulations, if the electric current is assumed to follow a completely uniform distribution with stream-lines filling a perfect cylinder under the nanocontact, then, $j_{\rho}=0$ (i.e., there is no in-plane current) and in virtue of Eq.\ \ref{eq:simplifZL}, the ZL is suppressed and the simulation shows no traces of it.

Here we consider an analytical calculation for the current distribution in a nanocontact STNO and study the effect of the ZL torques in different device configurations. We find that the ZL torque modifies the threshold currents of the magnetic modes and their effective sizes. Further, we analyze how to control the ZL torque effects by varying the current distribution through changes in the device fabrication (thickness of the used electrodes).

\section{Results}

\begin{figure}[ht]
	\includegraphics[width=1\columnwidth]{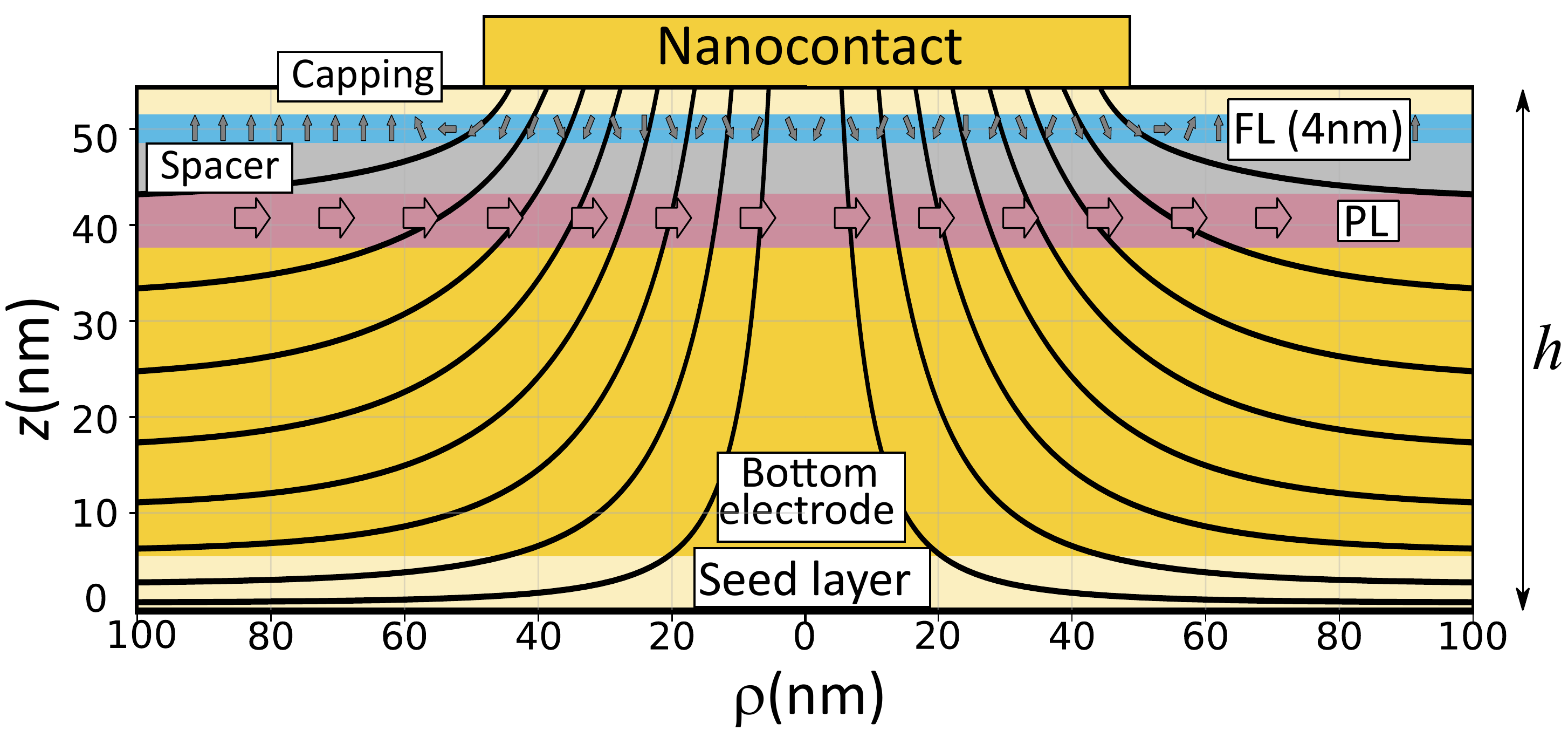}
	\caption{Schematic representation of a layer stack of a prototypical STNO with a FL and a PL magnetic layers plus a bottom electrode. Black lines show electric current streamlines in a vertical section (calculated with the model discussed in the Appendix).}
	\label{fig:stack}
\end{figure}

Although nanocontact SNTO are composed of several conducting layers, we consider here a single-layer device approximation to allow for analytical solutions of the charge current distribution (see, Appendix). Thus, the system becomes an infinite ohmic two dimensional layer with a given thickness, $h$, which accounts for both magnetic and non magnetic layers (see, Fig.\ \ref{fig:stack}). We first consider a magnetic FL of 4 nanometers of thickness in a total metallic stack of 54 (there is a non-magnetic capping layer of 2 nanometers above and another of 48 underneath accounting for a seed layer, an electrode, a PL and a spacer) as shown in Fig\ \ref{fig:stack}.

We now proceed to study the effects of the ZL torque on different spin wave excitations obtained in STNO with nanocontact geometry. We have simulated the evolution of a magnetic layer as a function of the polarized current density in a nanocontact of 50 nanometers of radius and $h=54$ nanometers for a layer with the equilibrium magnetization in the plane and with \textit{A)} an applied field perpendicular to the film plane, which results in an emission of spin waves, and \textit{B)} an applied field in the film plane, which creates localized bullets. We also studied the case \textit{C)} of a layer with perpendicular magnetic anisotropy (PMA) and an applied field perpendicular to the film plane, which originates droplet solitons.

\begin{figure}[ht]
\includegraphics[width=1\columnwidth]{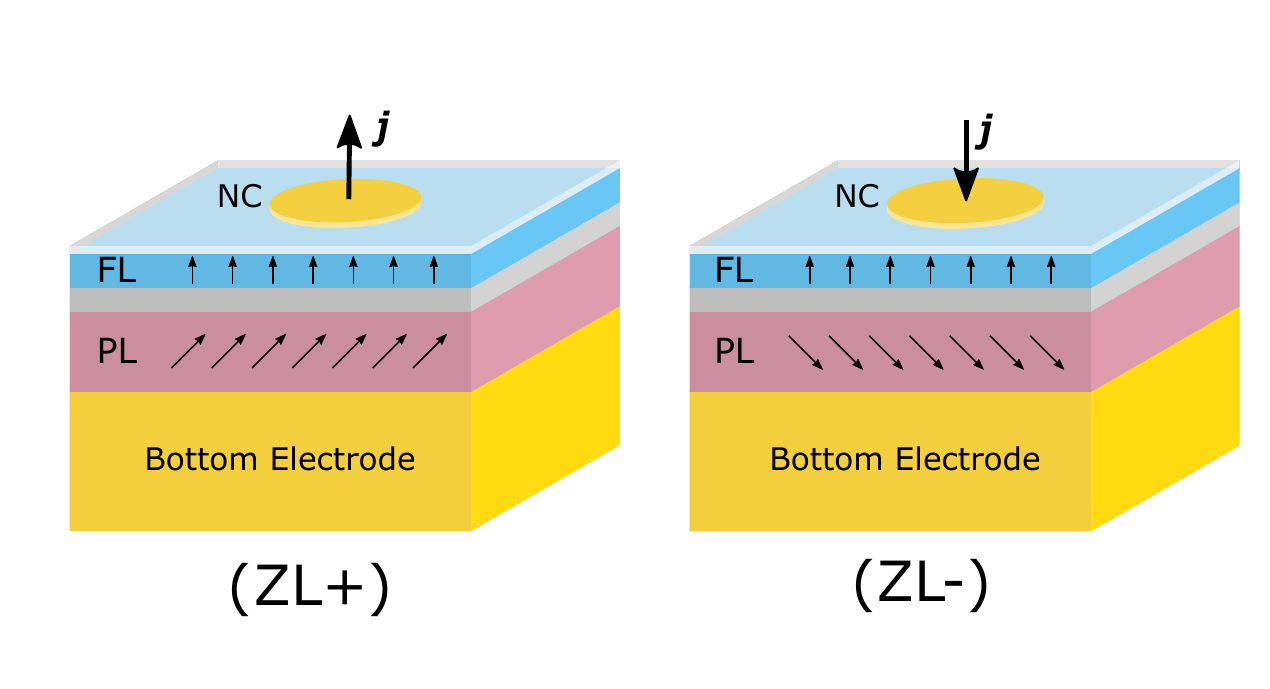}
\caption{\small{Schematic representation of STNO devices in two different configurations; the ZL$-$ and the ZL$+$ cases.}}
\label{fig:cases}
\end{figure}

In order to discern the effects of ZL torque from those of the LL and SL torques (in each type of excitation; \textit{A}, \textit{B}, and \textit{C}) we have considered three different configurations of STNO and compared the results obtained for each of them: \textit{i)} a reference case (no ZL), we have manually disabled the contribution of the ZL torque and $j_z>0$ (i.e.,\ we have removed the corresponding term from the equation of motion), \textit{ii)} a positive ZL case (ZL$+$) corresponding to a positive charge current, $j_z>0$ (it can be done experimentally having the PL magnetization with component along the FL magnetization), and  \textit{iii)} a negative ZL case (ZL$-$) corresponding to a negative charge current, $j_z<0$ (this could be the case when the PL magnetization is opposing the FL magnetization). Figure \ \ref{fig:cases} shows schematically the configuration of both FL and PL magnetization for the cases where the FL magnetization is out of the film plane. The LL and SL torques are identical in all three cases and only the ZL torque differs between them. This is so because the LL torque does not depend on $j_z$ nor on $(m_p)_z$ while the SL is proportional to both of them---thus, changing both signs leaves unchanged these torques. Contrarily, the ZL torque depends only on $j_\rho$, which has always the opposite sign of $j_z$ (as seen in Fig.\ \ref{fig:stack}). Therefore, the ZL torque in the ZL$+$ and ZL$-$ cases only differs in a sign. 

We performed micromagnetic simulations using the open-source MuMax$^3$ \cite{Vansteenkiste-mumax}. The parameters for the micromagnetic simulations are the following: saturation magnetization, $M_s=5\cdot 10^5$ A/m, damping constant $\alpha=0.03$, exchange stiffness $A=10^{-12}$J/m and for the case of PMA a uniaxial anisotropy constant $K_u= 2\cdot 10^5$ J/m$^3$. We assumed an electrical current with a spin polarization $p=0.5$ and an applied field of 0.8 T either in plane or out of the film plane.

We have computed current densities and associated Oersted fields using Eq.\ \ref{eq:Oe} from the Appendix. Notice that the distributions are identical for all considered cases (i.e., ZL$+$, no-ZL, ZL$-$ for \textit{A}, \textit{B}, \textit{C}) up to an overall sign and the intensity at the nanocontact, which is variable. Then, for each case, we have simulated the evolution of the FL magnetization with a slowly-increasing intensity of the applied current (in absolute value) at the nanocontact.


\subsection{Propagating spin waves}



When the FL is magnetized out of the film plane, pro\-pagating spin waves with a wave vector related to the inverse of the nanocontact radius can be generated, as described by Slonczewsky \cite{Slonczewski1999,Rippard2004}. We have studied the dependence of the spin wave frequency on the applied current at the nanocontact for each ZL case.

\begin{figure}[h!]
\includegraphics[width=1\columnwidth]{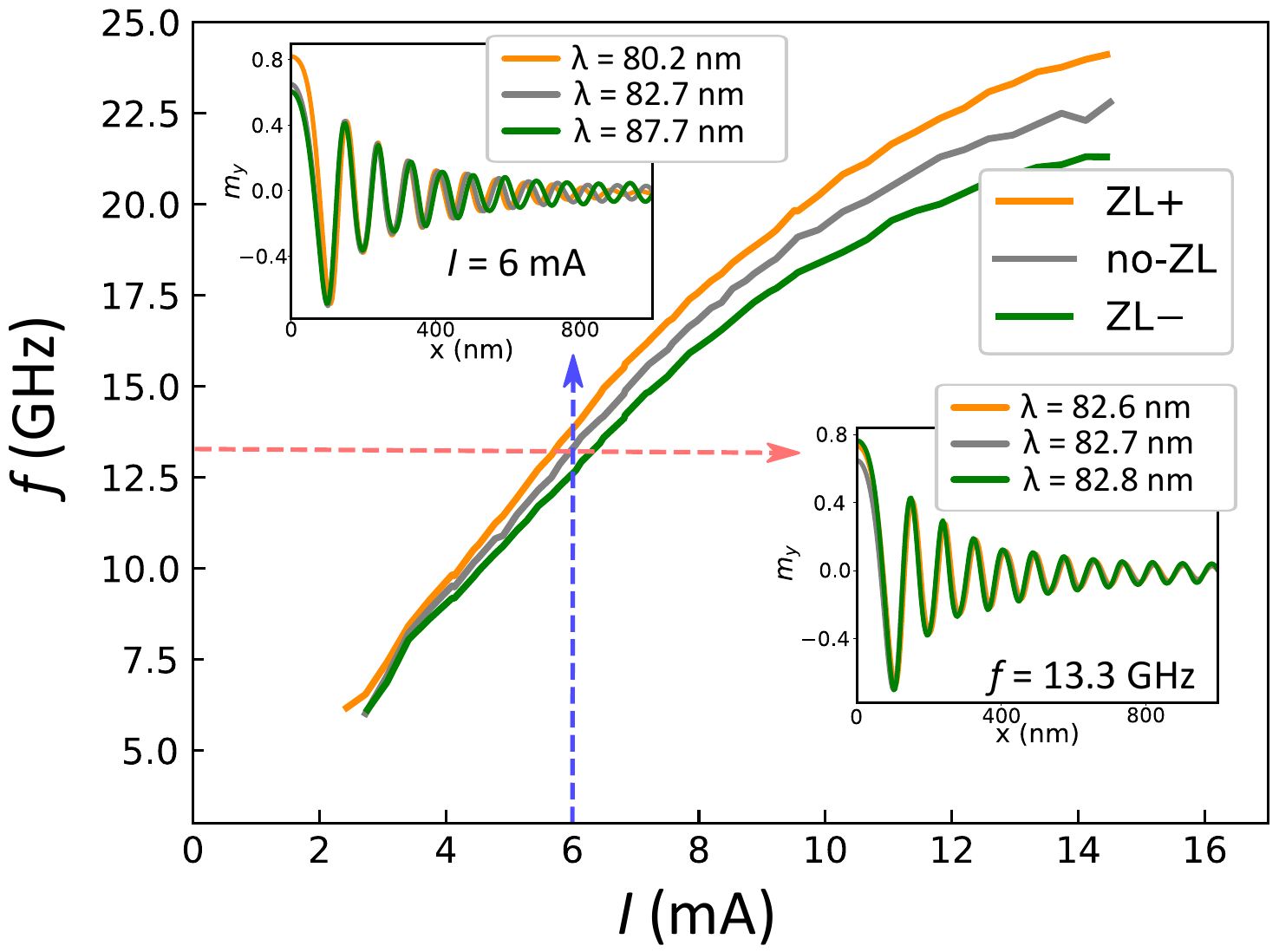}
\caption{\small{Spin wave frequency {\it vs} applied current at the nanocontact for current sweeps for different ZL cases. Insets show an in-plane component of the magnetization, $m_y$, in a central section for all ZL cases; in the upper inset at a fixed current ($I=6$ mA) and in the lower inset at a same frequency ($f=13.3\,$ GHz).}}
\label{fig:NoPMA}
\end{figure}

We can see in Fig.\ \ref{fig:NoPMA} a linear dependence of the oscillation frequency with the applied current in all three cases, in agreement with experimental results \cite{Mancoff2006,Madami2011} (we see also how the frequency saturates at large current values). We note that the spin-wave onset (i.e.,\ where it first appears a frequency peak during the current sweep) slightly differs in the three cases. The current threshold is, thus, affected by the ZL torque. Namely, the ZL$+$ case leads to a lower threshold whereas the ZL$-$ requires a higher current value, as compared to the no-ZL case. The whole frequency vs current curve is indeed affected by the ZL torque having different slopes.

We also analyze how the ZL torque affects spin-wave wavelengths. We have compared the $y$ component of the magnetization in a central section of the magnetic layer for each ZL case when they are either at the same current or at the same frequency (see insets in Fig.\ \ref{fig:NoPMA}). By averaging the wavelength of several shots within a period, we obtained that spin waves at a fixed current have different frequency and wavelength whereas spin waves at a fixed frequency have almost the same wavelength. The dispersion relation is thus barely affected by the ZL torque and, instead we could see the positive Zhang-Li torque as an additional force resulting in larger amplitude spin waves at the nanocontact---equivalent to a larger current value---, which sets a larger frequency excitation (with a lower wavelength).

\subsection{Localized bullets}



When a magnetic field is applied in-plane ($x$-axis in our case) a spin current excites a strongly nonlinear, self-localized solitonic bullet mode \cite{SlavinBullet,Consolo_prb2007,Bonetti2010}. Here to follow the evolution of the localized excitation over the current sweep for all ZL cases we plot the overall magnetization along the applied field direction within the simulated area normalized to the case where all film's magnetization points towards the applied field and the nanocontac region mangetization is completely reversed---this is an estimation of the spatial extension of the bullet mode.

\begin{figure}[h!]
\includegraphics[width=1\columnwidth]{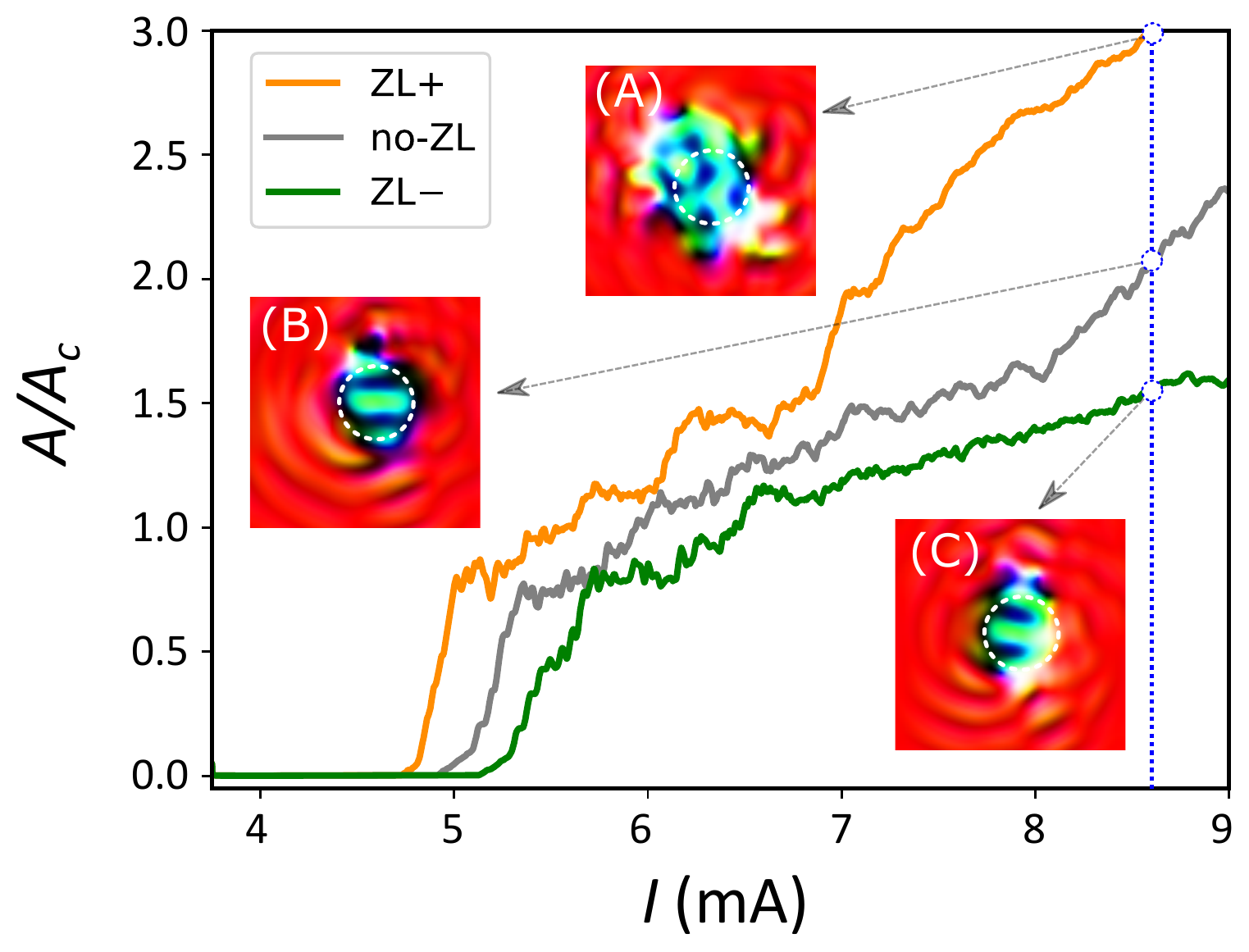}
\caption{\small{Effective area of the bullet excitation {\it vs} applied current for different ZL cases. Insets show $m_x$ of the FL for (A) ZL$+$, (B) no-ZL and (C) ZL$-$ at a same current of 8.75 mA.}}
\label{fig:inplane}
\end{figure}

Despite the oscillating nature of the bullet excitation, we can see in Fig.\ \ref{fig:inplane} that the ZL torque has qualitatively the same impact on generation current thresholds as it had for spin waves (i.e., ZL$+$ (ZL$-$) presents a lower (higher) threshold than the reference case, no-ZL).  In addition, difference in the effective area at higher currents presumably implies that bullets of different ZL cases differ significantly in size--being larger for ZL$+$ and smaller for the ZL$-$ compared to the reference case (See, insets of Fig.\ \ref{fig:inplane}).

\subsection{Magnetic droplet solitons}

When a magnetic layer has PMA, a spin current in a STNO can compensate damping and create droplet solitons \cite{Hoefer2010,Mohseni2013}. In this case the film's magnetization is out of the film plane and the droplet has its magnetization partially reversed and precessing. Here we also define an estimate for the spatial extension of the droplet as we did for bullet excitations.

\begin{figure}[h!]
\includegraphics[width=1\columnwidth]{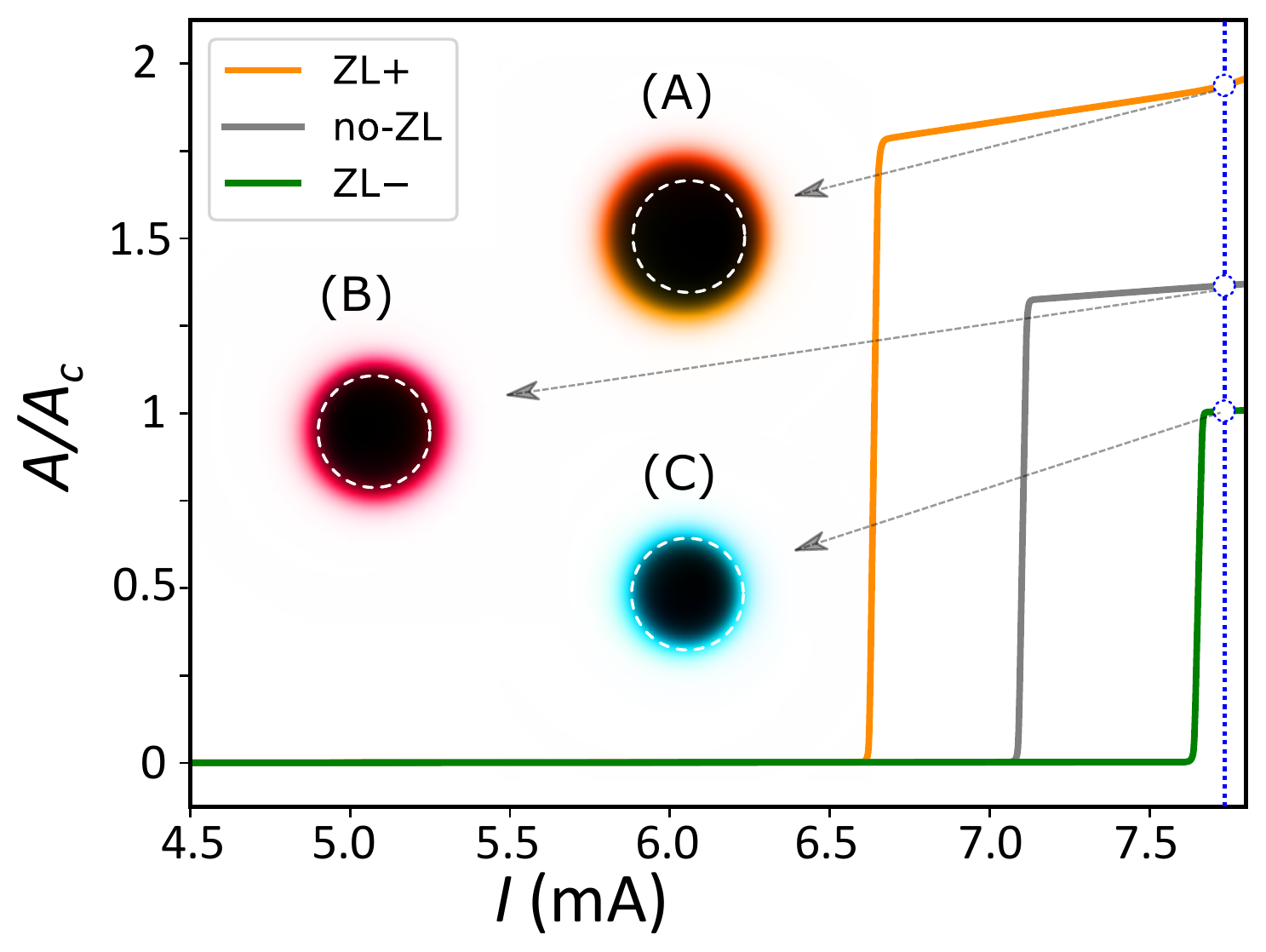}
\caption{\small{Effective area of droplet {\it vs} applied current at the nanocontact for different ZL cases. Insets show $m_z$ of the FL for the cases (A) ZL$+$, (B) no-ZL and (c) ZL$-$ at the same current ($I=7.75\,$ mA). The dashed white circle in them represents the nancocontact, whose diameter is 100 nm.}}
\label{fig:PMA}
\end{figure}

The dependence of droplet spatial extension on the applied current at the nanocontact is shown in Fig.\ \ref{fig:PMA}. Droplet nucleation occurs suddenly \cite{Jinting_2017}, so one can assign a well-defined value to threshold currents. Once again, the ZL$+$ (ZL$-$) threshold is lower (higher) than that of the reference case, and a similar effect happens to droplet size. Namely, once generated, ZL$+$ (ZL$-$) droplets are larger (smaller) than those of the reference case, as can be cross-checked with insets of Fig.\ \ref{fig:PMA}. This droplet enlargement was observed experimentally in X-ray microscopy images of droplets \cite{AkermanXMCD}.

\section{Controlling the ZL torque}
We have compared the evolution of spin wave excitations in STNO as a function of the applied current for positive and negative ZL torques. Now we study how to control and modulate the strength of the ZL torque in an STNO focusing on the droplet soliton case.

In STNO, the ZL torque is proportional to the in-plane current density (Eq.\ \ref{eq:simplifZL}), which can be controlled by the ratio between device's thickness, $h$, and the nanocontact diameter. A thick electrode would produce an almost uniform current distribution in the magnetic layer (i.e.,\ with a small in-plane component) while a thin electrode would lead to more spread-out currents in the magnetic layer (as that of Fig.\ \ref{fig:stack}), with a saturation at the $h\rightarrow\infty$ limit. Thus, by changing the device's electrode thickness (or the nanocontac diameter) we would be able to modulate the ZL torque.

We have simulated current sweeps for different electrode thicknesses keeping the stack configuration constant as described in Fig.\ \ref{fig:stack} for the no-ZL, ZL$+$ and ZL$-$ cases. We obtained curves as in Fig.\ \ref{fig:PMA} for each thickness, $h$, from which we have kept representative values of the threshold current and spatial extension. We summarize our results in Fig.\ \ref{fig:thick}. The droplet spatial size is increased by ZL$+$ and decreased by ZL$-$, as compared to the no-ZL case for all values of $h$ and current thresholds are lower at ZL$+$, and higher at ZL$-$, as discussed in the previous section. 

We can also see that these effects are more important at small values of $h$ because the in-plane current density is higher. We also notice that a saturation value is obtained for thicknesses between 75 and 100 nm, which correspond to current density distributions similar to the case $h\rightarrow\infty$. We note here that the current threshold in the no-ZL case  also depends on $h$. This fact is due to charge conservation; when $h$ is thinner, the current distribution spreads out and $j_\rho$ increases at expenses of $j_z$. Thus, $|j_z|$ decreases, reducing the SL torque and increasing the necessary applied current to reverse {$\mb{m}$ and generate the droplet. It is for this reason that threshold current values increase with thinner $h$ in all cases. But the ZL$+$ torque counteracts this fact and leads to smaller threshold increments whereas the ZL$-$ torque further increases them.

\begin{figure}[h!]
\includegraphics[width=1\columnwidth]{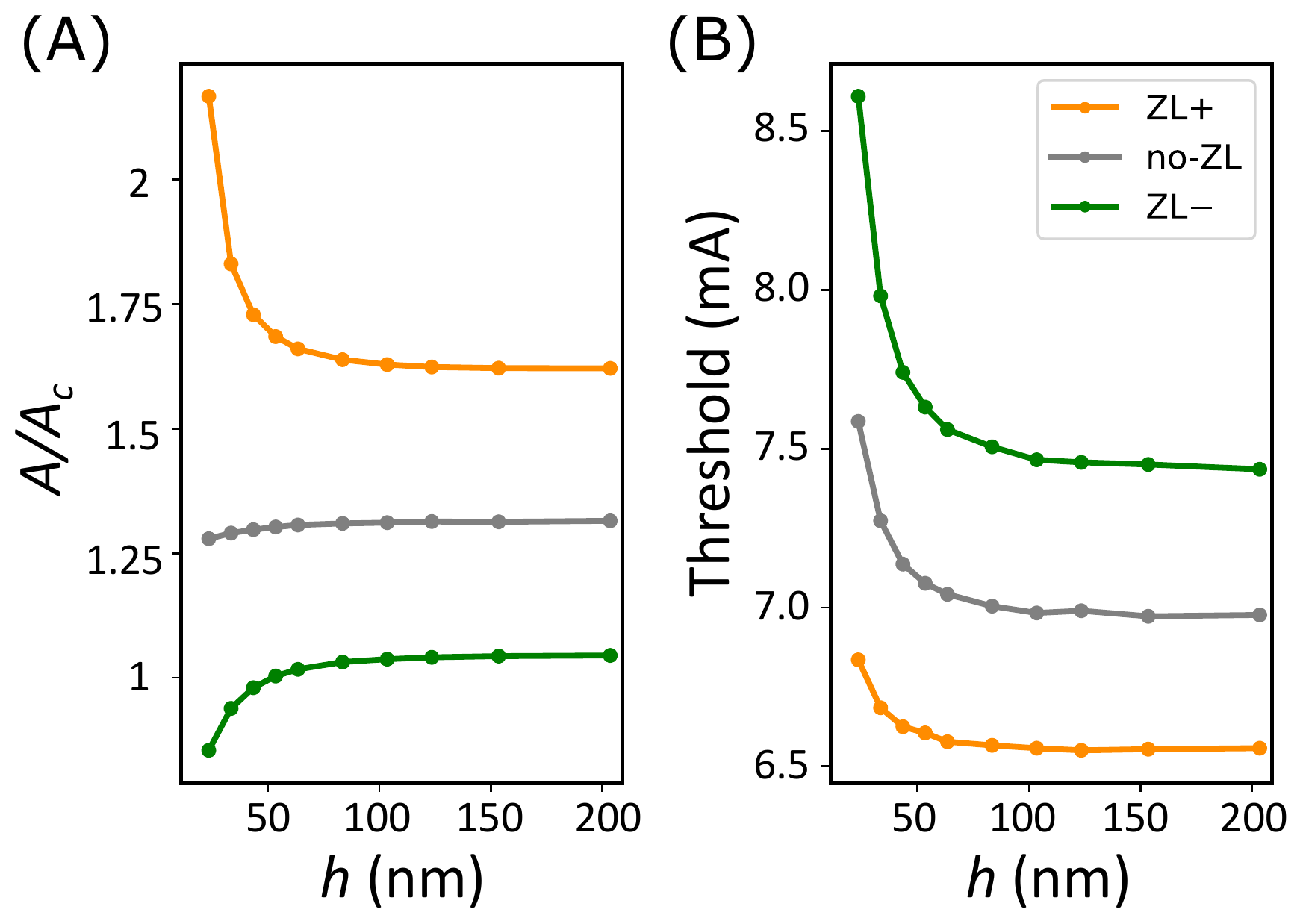}
\caption{\small{In (A) effective area of droplet at a representative current of $I=7.6\,$mA {\it vs} layer thickness, $h$. In (B) current threshold values {\it vs} layer thickness, $h$.}}
\label{fig:thick}
\end{figure}


\section{Conclusions}

We have studied the effect of Zhang-Li torques in magnetic excitations produced in STNOs with a nanocontact geometry. Our simulations show that the ZL torque modifies the threshold currents of the magnetic modes and their effective sizes. Given that only in-plane currents produce sizable Zhang-Li torques in magnetic thin films, we showed that changing the ratio between nanocontact size and layer thicknesses allows for a broad control of the Zhang-Li torque and its effects. We found that using electrodes thinner than 50-75 nm (for a nanocontact of 50 nm fo radius) produce changes of current threshold and effective size of excitations up to 20 \%. Zhang-Li torque effects might also be relevant in determining spatial extensions in magnetic excitations from spin-hall nano oscilators \cite{simulation_shno} where current paths are less symmetric than STNO with nanocontact geometries.



\section{ACKNOWLEDGMENTS}
FM acknowledges support from the RyC through Grant No. RYC-2014-16515. JM and FM acknowledge funding from MINECO through Grant No. MAT2015-69144-P.

\section{Appendix: Electric current and Oersted fields}
Although nanocontact SNTO are composed of several conducting layers, we will consider here a single-layer device approximation to allow for analytical solutions. Additionally, we will neglect the Hall effect produced by the applied magnetic field. Thus, the system becomes an infinite ohmic layer set between two insulators, the top one counting with an electrical nanocontact with cylindrical symmetry (Fig.\ \ref{fig:system}). 
\begin{figure}[h!]
	\centering
	\includegraphics[width=1\columnwidth]{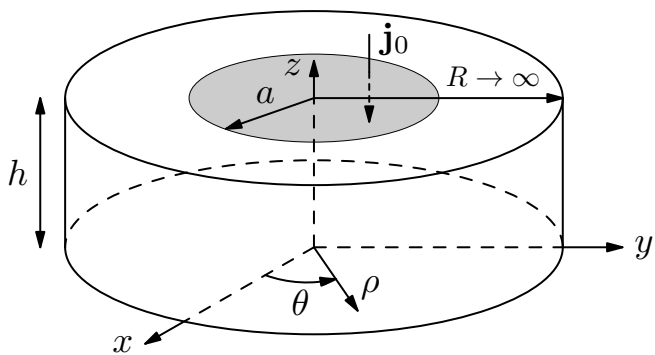}
	\caption{\small{Simplified geometry of a STNO.}}
	\label{fig:system}
\end{figure}

In ohmic materials, the equation that describes stationary electric currents is Laplace's equation for the electric potential $V$; $\nabla^2 V = 0$, where the electric current density, $\mb{j}$, is related to the electric potential through the resistivity $\varrho$ as
\begin{equation}
\mb{j} = -\frac{1}{\varrho}\bm{\nabla}V.
\label{eq:j}
\end{equation}

Then, by working in cylindrical coordinates, using the separation of variables method and assuming $V$ does not depend on $\theta$, one gets the general solution
$$V(\rho,z) = (C_1\sinh(kz) + C_2\cosh(kz))(J_0 (k\rho) + C_3N_0 (k\rho)),$$
where $C_1, C_2, C_3$ and $k$ are constants to be determined by boundary conditions and $J_0$ and $N_0$ are the 0th order Bessel functions of first and second kind respectively.

The boundary conditions (BCs) of this ideal system correspond to those of a perfect electrical nanocontact and perfect insulators, namely
\begin{enumerate}
	\itemsep-0.3em
	\item[(i)] $\quad V(\rho,z)<\infty \quad \rho\rightarrow 0$
	\item[(ii)] $\quad j_z(\rho,z=0)= 0 $
	\item[(iii)] $\quad V(\rho,z)=0 \quad \rho\rightarrow \infty$
	\item[(iv)] $\quad j_z(\rho,z=h)= j_0 \,\theta(a-\rho)$.
\end{enumerate}

BC (i) results from requiring the electric potential to be finite everywhere, especially at the center of the system. This BC accounts for the fact that energy cannot diverge and it implies $C_3=0$.

BC (ii) sets the zero of the perpendicular current density at the bottom of the device. Applying this BC leads to $C_1=0$.

BC (iii) sets the zero of $V$ at infinity in the radial direction, implicitly stating that the radial width of the system is very large compared to its height. In a finite system, this BC would restrict the possible values of $k$ by a relation with the discrete roots of Bessel functions. However, since the boundary is sent to infinity, such a restriction does not appear and $k$ becomes a continuous variable. The most general solution is then the sum (actually an integration) for all possible values of $k$; and $C_2$ becomes $C_2(k)$.

Finally, BC (iv) states that a given uniform current $j_0$ enters the cylinder through a nanocontact of radius $a$ placed at $z=h$. By using Eq.\ (\ref{eq:j}) and the orthogonality relation of Bessel functions, one gets an expression for $C_2(k)$ that leads to the solution
\begin{equation}
V(\rho,z) = -\varrho \, j_0\, a \int_{0}^\infty \frac{\cosh(kz)J_1 (k a)J_0 (k\rho)}{k\sinh(k h)} \, dk.
\label{eq:pot}
\end{equation}

To get the electric current density $\mb{j}$ one just needs to plug Eq.\ (\ref{eq:pot}) into Eq.\ (\ref{eq:j}). Note that $j_{\theta}=0$ because $V$ does not depend on $\theta$. Fig.\ \ref{fig:stack} shows electric current field lines for a case with $a=50\,$nm, $h=54\,$nm. Note how the distribution is far from the uniform case and that the FL experiences $j_\rho\neq 0$. 

According to the Biot-Savart law (BS law), electric currents generate magnetic fields that are usually called Oersted fields, so let us compute those generated by our electric current distribution. Due to the symmetry of the system, the magnetic field will only have a nonzero component in the azimuthal direction. Then, by using Amp\`ere's law one gets a simple expression for the generated magnetic field.
\begin{equation}
\mb{B} = \frac{\mu_0}{\rho} \int_{0}^{\rho}j_z \,\rho' \, d\rho' \,\, \mb{a}_{\theta}.
\label{eq:Oe}
\end{equation}

This expression only depends on the $z$ component of the inner current, and it is computed by a simple integration rather than the 3-variable integration of the BS law. Therefore, the current density and the magnetic field need not be computed outside the FL saving, thus, much computing time. Both expressions (Eq.\ \ref{eq:pot}---together with Eq.\ \ref{eq:j}---for the electric current density and Eq.\ \ref{eq:Oe} for the Oersted fields) are now simple enough to be quickly evaluated using numerical integration methods.


\begin{thebibliography}{33}%
	\makeatletter
	\providecommand \@ifxundefined [1]{%
		\@ifx{#1\undefined}
	}%
	\providecommand \@ifnum [1]{%
		\ifnum #1\expandafter \@firstoftwo
		\else \expandafter \@secondoftwo
		\fi
	}%
	\providecommand \@ifx [1]{%
		\ifx #1\expandafter \@firstoftwo
		\else \expandafter \@secondoftwo
		\fi
	}%
	\providecommand \natexlab [1]{#1}%
	\providecommand \enquote  [1]{``#1''}%
	\providecommand \bibnamefont  [1]{#1}%
	\providecommand \bibfnamefont [1]{#1}%
	\providecommand \citenamefont [1]{#1}%
	\providecommand \href@noop [0]{\@secondoftwo}%
	\providecommand \href [0]{\begingroup \@sanitize@url \@href}%
	\providecommand \@href[1]{\@@startlink{#1}\@@href}%
	\providecommand \@@href[1]{\endgroup#1\@@endlink}%
	\providecommand \@sanitize@url [0]{\catcode `\\12\catcode `\$12\catcode
		`\&12\catcode `\#12\catcode `\^12\catcode `\_12\catcode `\%12\relax}%
	\providecommand \@@startlink[1]{}%
	\providecommand \@@endlink[0]{}%
	\providecommand \url  [0]{\begingroup\@sanitize@url \@url }%
	\providecommand \@url [1]{\endgroup\@href {#1}{\urlprefix }}%
	\providecommand \urlprefix  [0]{URL }%
	\providecommand \Eprint [0]{\href }%
	\providecommand \doibase [0]{http://dx.doi.org/}%
	\providecommand \selectlanguage [0]{\@gobble}%
	\providecommand \bibinfo  [0]{\@secondoftwo}%
	\providecommand \bibfield  [0]{\@secondoftwo}%
	\providecommand \translation [1]{[#1]}%
	\providecommand \BibitemOpen [0]{}%
	\providecommand \bibitemStop [0]{}%
	\providecommand \bibitemNoStop [0]{.\EOS\space}%
	\providecommand \EOS [0]{\spacefactor3000\relax}%
	\providecommand \BibitemShut  [1]{\csname bibitem#1\endcsname}%
	\let\auto@bib@innerbib\@empty
	\bibitem [{\citenamefont {Tsoi}\ \emph {et~al.}(1998)\citenamefont {Tsoi},
		\citenamefont {Jansen}, \citenamefont {Bass}, \citenamefont {Chiang},
		\citenamefont {Seck}, \citenamefont {Tsoi},\ and\ \citenamefont
		{Wyder}}]{Tsoi1998}%
	\BibitemOpen
	\bibfield  {author} {\bibinfo {author} {\bibfnamefont {M.}~\bibnamefont
			{Tsoi}}, \bibinfo {author} {\bibfnamefont {A.~G.~M.}\ \bibnamefont {Jansen}},
		\bibinfo {author} {\bibfnamefont {J.}~\bibnamefont {Bass}}, \bibinfo {author}
		{\bibfnamefont {W.-C.}\ \bibnamefont {Chiang}}, \bibinfo {author}
		{\bibfnamefont {M.}~\bibnamefont {Seck}}, \bibinfo {author} {\bibfnamefont
			{V.}~\bibnamefont {Tsoi}}, \ and\ \bibinfo {author} {\bibfnamefont
			{P.}~\bibnamefont {Wyder}},\ }\href {\doibase 10.1103/PhysRevLett.80.4281}
	{\bibfield  {journal} {\bibinfo  {journal} {Phys. Rev. Lett.}\ }\textbf
		{\bibinfo {volume} {80}},\ \bibinfo {pages} {4281} (\bibinfo {year}
		{1998})}\BibitemShut {NoStop}%
	\bibitem [{\citenamefont {Slonczewski}(1999)}]{Slonczewski1999}%
	\BibitemOpen
	\bibfield  {author} {\bibinfo {author} {\bibfnamefont {J.~C.}\ \bibnamefont
			{Slonczewski}},\ }\href {\doibase 10.1016/S0304-8853(99)00043-8} {\bibfield
		{journal} {\bibinfo  {journal} {J. Magn. Magn. Mater.}\ }\textbf {\bibinfo
			{volume} {195}},\ \bibinfo {pages} {L261} (\bibinfo {year}
		{1999})}\BibitemShut {NoStop}%
	\bibitem [{\citenamefont {Kiselev}\ \emph {et~al.}(2003)\citenamefont
		{Kiselev}, \citenamefont {Sankey}, \citenamefont {Krivorotov}, \citenamefont
		{Emley}, \citenamefont {Schoelkopf}, \citenamefont {Buhrman},\ and\
		\citenamefont {Ralph}}]{Kiselev2003}%
	\BibitemOpen
	\bibfield  {author} {\bibinfo {author} {\bibfnamefont {S.~I.}\ \bibnamefont
			{Kiselev}}, \bibinfo {author} {\bibfnamefont {J.~C.}\ \bibnamefont {Sankey}},
		\bibinfo {author} {\bibfnamefont {I.~N.}\ \bibnamefont {Krivorotov}},
		\bibinfo {author} {\bibfnamefont {N.~C.}\ \bibnamefont {Emley}}, \bibinfo
		{author} {\bibfnamefont {R.~J.}\ \bibnamefont {Schoelkopf}}, \bibinfo
		{author} {\bibfnamefont {R.~A.}\ \bibnamefont {Buhrman}}, \ and\ \bibinfo
		{author} {\bibfnamefont {D.~C.}\ \bibnamefont {Ralph}},\ }\href {\doibase
		10.1038/nature01967} {\bibfield  {journal} {\bibinfo  {journal} {Nature}\
		}\textbf {\bibinfo {volume} {425}},\ \bibinfo {pages} {380} (\bibinfo {year}
		{2003})}\BibitemShut {NoStop}%
	\bibitem [{\citenamefont {Rippard}\ \emph {et~al.}(2004)\citenamefont
		{Rippard}, \citenamefont {Pufall}, \citenamefont {Kaka}, \citenamefont
		{Russek},\ and\ \citenamefont {Silva}}]{Rippard2004}%
	\BibitemOpen
	\bibfield  {author} {\bibinfo {author} {\bibfnamefont {W.}~\bibnamefont
			{Rippard}}, \bibinfo {author} {\bibfnamefont {M.}~\bibnamefont {Pufall}},
		\bibinfo {author} {\bibfnamefont {S.}~\bibnamefont {Kaka}}, \bibinfo {author}
		{\bibfnamefont {S.}~\bibnamefont {Russek}}, \ and\ \bibinfo {author}
		{\bibfnamefont {T.}~\bibnamefont {Silva}},\ }\href {\doibase
		10.1103/PhysRevLett.92.027201} {\bibfield  {journal} {\bibinfo  {journal}
			{Phys. Rev. Lett.}\ }\textbf {\bibinfo {volume} {92}},\ \bibinfo {pages}
		{027201} (\bibinfo {year} {2004})}\BibitemShut {NoStop}%
	\bibitem [{\citenamefont {Heyne}\ \emph {et~al.}(2010)\citenamefont {Heyne},
		\citenamefont {Rhensius}, \citenamefont {Ilgaz}, \citenamefont {Bisig},
		\citenamefont {R\"udiger}, \citenamefont {Kl\"aui}, \citenamefont {Joly},
		\citenamefont {Nolting}, \citenamefont {Heyderman}, \citenamefont {Thiele},\
		and\ \citenamefont {Kronast}}]{Vortex_heyne}%
	\BibitemOpen
	\bibfield  {author} {\bibinfo {author} {\bibfnamefont {L.}~\bibnamefont
			{Heyne}}, \bibinfo {author} {\bibfnamefont {J.}~\bibnamefont {Rhensius}},
		\bibinfo {author} {\bibfnamefont {D.}~\bibnamefont {Ilgaz}}, \bibinfo
		{author} {\bibfnamefont {A.}~\bibnamefont {Bisig}}, \bibinfo {author}
		{\bibfnamefont {U.}~\bibnamefont {R\"udiger}}, \bibinfo {author}
		{\bibfnamefont {M.}~\bibnamefont {Kl\"aui}}, \bibinfo {author} {\bibfnamefont
			{L.}~\bibnamefont {Joly}}, \bibinfo {author} {\bibfnamefont {F.}~\bibnamefont
			{Nolting}}, \bibinfo {author} {\bibfnamefont {L.~J.}\ \bibnamefont
			{Heyderman}}, \bibinfo {author} {\bibfnamefont {J.~U.}\ \bibnamefont
			{Thiele}}, \ and\ \bibinfo {author} {\bibfnamefont {F.}~\bibnamefont
			{Kronast}},\ }\href {\doibase 10.1103/PhysRevLett.105.187203} {\bibfield
		{journal} {\bibinfo  {journal} {Phys. Rev. Lett.}\ }\textbf {\bibinfo
			{volume} {105}},\ \bibinfo {pages} {187203} (\bibinfo {year}
		{2010})}\BibitemShut {NoStop}%
	\bibitem [{\citenamefont {Pribiag}\ \emph {et~al.}(2007)\citenamefont
		{Pribiag}, \citenamefont {Krivorotov}, \citenamefont {Fuchs}, \citenamefont
		{Braganca}, \citenamefont {Ozatay}, \citenamefont {Sankey}, \citenamefont
		{Ralph},\ and\ \citenamefont {Buhrman}}]{Vortex_nature}%
	\BibitemOpen
	\bibfield  {author} {\bibinfo {author} {\bibfnamefont {V.~S.}\ \bibnamefont
			{Pribiag}}, \bibinfo {author} {\bibfnamefont {I.~N.}\ \bibnamefont
			{Krivorotov}}, \bibinfo {author} {\bibfnamefont {G.~D.}\ \bibnamefont
			{Fuchs}}, \bibinfo {author} {\bibfnamefont {P.~M.}\ \bibnamefont {Braganca}},
		\bibinfo {author} {\bibfnamefont {O.}~\bibnamefont {Ozatay}}, \bibinfo
		{author} {\bibfnamefont {J.~C.}\ \bibnamefont {Sankey}}, \bibinfo {author}
		{\bibfnamefont {D.~C.}\ \bibnamefont {Ralph}}, \ and\ \bibinfo {author}
		{\bibfnamefont {R.~A.}\ \bibnamefont {Buhrman}},\ }\href {\doibase
		10.1038/nphys619} {\bibfield  {journal} {\bibinfo  {journal} {Nature
				Physics}\ }\textbf {\bibinfo {volume} {3}},\ \bibinfo {pages} {498} (\bibinfo
		{year} {2007})}\BibitemShut {NoStop}%
	\bibitem [{\citenamefont {Slavin}\ and\ \citenamefont
		{Tiberkevich}(2005)}]{SlavinBullet}%
	\BibitemOpen
	\bibfield  {author} {\bibinfo {author} {\bibfnamefont {A.}~\bibnamefont
			{Slavin}}\ and\ \bibinfo {author} {\bibfnamefont {V.}~\bibnamefont
			{Tiberkevich}},\ }\href {\doibase 10.1103/PhysRevLett.95.237201} {\bibfield
		{journal} {\bibinfo  {journal} {Phys. Rev. Lett.}\ }\textbf {\bibinfo
			{volume} {95}},\ \bibinfo {pages} {237201} (\bibinfo {year}
		{2005})}\BibitemShut {NoStop}%
	\bibitem [{\citenamefont {Madami}\ \emph {et~al.}(2011)\citenamefont {Madami},
		\citenamefont {Bonetti}, \citenamefont {Consolo}, \citenamefont {Tacchi},
		\citenamefont {Carlotti}, \citenamefont {Gubbiotti}, \citenamefont {Mancoff},
		\citenamefont {Yar},\ and\ \citenamefont {Akerman}}]{Madami2011}%
	\BibitemOpen
	\bibfield  {author} {\bibinfo {author} {\bibfnamefont {M.}~\bibnamefont
			{Madami}}, \bibinfo {author} {\bibfnamefont {S.}~\bibnamefont {Bonetti}},
		\bibinfo {author} {\bibfnamefont {G.}~\bibnamefont {Consolo}}, \bibinfo
		{author} {\bibfnamefont {S.}~\bibnamefont {Tacchi}}, \bibinfo {author}
		{\bibfnamefont {G.}~\bibnamefont {Carlotti}}, \bibinfo {author}
		{\bibfnamefont {G.}~\bibnamefont {Gubbiotti}}, \bibinfo {author}
		{\bibfnamefont {F.~B.}\ \bibnamefont {Mancoff}}, \bibinfo {author}
		{\bibfnamefont {M.~A.}\ \bibnamefont {Yar}}, \ and\ \bibinfo {author}
		{\bibfnamefont {J.}~\bibnamefont {Akerman}},\ }\href {\doibase
		doi:10.1038/nnano.2011.140} {\bibfield  {journal} {\bibinfo  {journal} {Nat.
				Nanotechnol.}\ }\textbf {\bibinfo {volume} {6}},\ \bibinfo {pages} {635}
		(\bibinfo {year} {2011})}\BibitemShut {NoStop}%
	\bibitem [{\citenamefont {Mohseni}\ \emph {et~al.}(2013)\citenamefont
		{Mohseni}, \citenamefont {Sani}, \citenamefont {Persson}, \citenamefont
		{Nguyen}, \citenamefont {Chung}, \citenamefont {Pogoryelov}, \citenamefont
		{Muduli}, \citenamefont {Iacocca}, \citenamefont {Eklund}, \citenamefont
		{Dumas}, \citenamefont {Bonetti}, \citenamefont {Deac}, \citenamefont
		{Hoefer},\ and\ \citenamefont {{\AA}kerman}}]{Mohseni2013}%
	\BibitemOpen
	\bibfield  {author} {\bibinfo {author} {\bibfnamefont {S.~M.}\ \bibnamefont
			{Mohseni}}, \bibinfo {author} {\bibfnamefont {S.~R.}\ \bibnamefont {Sani}},
		\bibinfo {author} {\bibfnamefont {J.}~\bibnamefont {Persson}}, \bibinfo
		{author} {\bibfnamefont {T.~N.~A.}\ \bibnamefont {Nguyen}}, \bibinfo {author}
		{\bibfnamefont {S.}~\bibnamefont {Chung}}, \bibinfo {author} {\bibfnamefont
			{Y.}~\bibnamefont {Pogoryelov}}, \bibinfo {author} {\bibfnamefont {P.~K.}\
			\bibnamefont {Muduli}}, \bibinfo {author} {\bibfnamefont {E.}~\bibnamefont
			{Iacocca}}, \bibinfo {author} {\bibfnamefont {A.}~\bibnamefont {Eklund}},
		\bibinfo {author} {\bibfnamefont {R.~K.}\ \bibnamefont {Dumas}}, \bibinfo
		{author} {\bibfnamefont {S.}~\bibnamefont {Bonetti}}, \bibinfo {author}
		{\bibfnamefont {A.}~\bibnamefont {Deac}}, \bibinfo {author} {\bibfnamefont
			{M.~A.}\ \bibnamefont {Hoefer}}, \ and\ \bibinfo {author} {\bibfnamefont
			{J.}~\bibnamefont {{\AA}kerman}},\ }\href {\doibase 10.1126/science.1230155}
	{\bibfield  {journal} {\bibinfo  {journal} {Science}\ }\textbf {\bibinfo
			{volume} {339}},\ \bibinfo {pages} {1295} (\bibinfo {year}
		{2013})}\BibitemShut {NoStop}%
	\bibitem [{\citenamefont {Maci\`{a}}\ \emph {et~al.}(2014)\citenamefont
		{Maci\`{a}}, \citenamefont {Backes},\ and\ \citenamefont {Kent}}]{Macia2014}%
	\BibitemOpen
	\bibfield  {author} {\bibinfo {author} {\bibfnamefont {F.}~\bibnamefont
			{Maci\`{a}}}, \bibinfo {author} {\bibfnamefont {D.}~\bibnamefont {Backes}}, \
		and\ \bibinfo {author} {\bibfnamefont {A.~D.}\ \bibnamefont {Kent}},\ }\href
	{\doibase 10.1038/nnano.2014.255} {\bibfield  {journal} {\bibinfo  {journal}
			{Nat. Nanotechnol.}\ }\textbf {\bibinfo {volume} {9}},\ \bibinfo {pages}
		{992} (\bibinfo {year} {2014})}\BibitemShut {NoStop}%
	\bibitem [{\citenamefont {Demidov}\ \emph {et~al.}(2010)\citenamefont
		{Demidov}, \citenamefont {Urazhdin},\ and\ \citenamefont
		{Demokritov}}]{Demidov2010}%
	\BibitemOpen
	\bibfield  {author} {\bibinfo {author} {\bibfnamefont {V.~E.}\ \bibnamefont
			{Demidov}}, \bibinfo {author} {\bibfnamefont {S.}~\bibnamefont {Urazhdin}}, \
		and\ \bibinfo {author} {\bibfnamefont {S.~O.}\ \bibnamefont {Demokritov}},\
	}\href {\doibase doi:10.1038/nmat2882} {\bibfield  {journal} {\bibinfo
			{journal} {Nat. Mater.}\ }\textbf {\bibinfo {volume} {9}},\ \bibinfo {pages}
		{984} (\bibinfo {year} {2010})}\BibitemShut {NoStop}%
	\bibitem [{\citenamefont {Bonetti}\ \emph {et~al.}(2015)\citenamefont
		{Bonetti}, \citenamefont {Kukreja}, \citenamefont {Chen}, \citenamefont
		{Maci\`{a}}, \citenamefont {Hern\`{a}ndez}, \citenamefont {Eklund},
		\citenamefont {Backes}, \citenamefont {Frisch}, \citenamefont {Katine},
		\citenamefont {Malm}, \citenamefont {Urazhdin}, \citenamefont {Kent},
		\citenamefont {St\"{o}hr}, \citenamefont {Ohldag},\ and\ \citenamefont
		{D\"{u}rr}}]{Bonetti2015}%
	\BibitemOpen
	\bibfield  {author} {\bibinfo {author} {\bibfnamefont {S.}~\bibnamefont
			{Bonetti}}, \bibinfo {author} {\bibfnamefont {R.}~\bibnamefont {Kukreja}},
		\bibinfo {author} {\bibfnamefont {Z.}~\bibnamefont {Chen}}, \bibinfo {author}
		{\bibfnamefont {F.}~\bibnamefont {Maci\`{a}}}, \bibinfo {author}
		{\bibfnamefont {J.~M.}\ \bibnamefont {Hern\`{a}ndez}}, \bibinfo {author}
		{\bibfnamefont {A.}~\bibnamefont {Eklund}}, \bibinfo {author} {\bibfnamefont
			{D.}~\bibnamefont {Backes}}, \bibinfo {author} {\bibfnamefont
			{J.}~\bibnamefont {Frisch}}, \bibinfo {author} {\bibfnamefont
			{J.}~\bibnamefont {Katine}}, \bibinfo {author} {\bibfnamefont
			{G.}~\bibnamefont {Malm}}, \bibinfo {author} {\bibfnamefont {S.}~\bibnamefont
			{Urazhdin}}, \bibinfo {author} {\bibfnamefont {A.~D.}\ \bibnamefont {Kent}},
		\bibinfo {author} {\bibfnamefont {J.}~\bibnamefont {St\"{o}hr}}, \bibinfo
		{author} {\bibfnamefont {H.}~\bibnamefont {Ohldag}}, \ and\ \bibinfo {author}
		{\bibfnamefont {H.~A.}\ \bibnamefont {D\"{u}rr}},\ }\href {\doibase
		doi:10.1038/ncomms9889} {\bibfield  {journal} {\bibinfo  {journal} {Nat.
				Commun.}\ }\textbf {\bibinfo {volume} {6}},\ \bibinfo {pages} {8889}
		(\bibinfo {year} {2015})}\BibitemShut {NoStop}%
	\bibitem [{\citenamefont {Slonczewski}(1996)}]{Slonczewski1996}%
	\BibitemOpen
	\bibfield  {author} {\bibinfo {author} {\bibfnamefont {J.~C.}\ \bibnamefont
			{Slonczewski}},\ }\href {\doibase doi.org/10.1016/0304-8853(96)00062-5}
	{\bibfield  {journal} {\bibinfo  {journal} {J. Magn. Magn. Mater.}\ }\textbf
		{\bibinfo {volume} {159}},\ \bibinfo {pages} {L1} (\bibinfo {year}
		{1996})}\BibitemShut {NoStop}%
	\bibitem [{\citenamefont {Berger}(1996)}]{Berger1996}%
	\BibitemOpen
	\bibfield  {author} {\bibinfo {author} {\bibfnamefont {L.}~\bibnamefont
			{Berger}},\ }\href {\doibase 10.1103/PhysRevB.54.9353} {\bibfield  {journal}
		{\bibinfo  {journal} {Phys. Rev. B}\ }\textbf {\bibinfo {volume} {54}},\
		\bibinfo {pages} {9353} (\bibinfo {year} {1996})}\BibitemShut {NoStop}%
	\bibitem [{\citenamefont {Kato}\ \emph {et~al.}(2004)\citenamefont {Kato},
		\citenamefont {Myers}, \citenamefont {Gossard},\ and\ \citenamefont
		{Awschalom}}]{Kato1910_SHE}%
	\BibitemOpen
	\bibfield  {author} {\bibinfo {author} {\bibfnamefont {Y.~K.}\ \bibnamefont
			{Kato}}, \bibinfo {author} {\bibfnamefont {R.~C.}\ \bibnamefont {Myers}},
		\bibinfo {author} {\bibfnamefont {A.~C.}\ \bibnamefont {Gossard}}, \ and\
		\bibinfo {author} {\bibfnamefont {D.~D.}\ \bibnamefont {Awschalom}},\ }\href
	{\doibase 10.1126/science.1105514} {\bibfield  {journal} {\bibinfo  {journal}
			{Science}\ }\textbf {\bibinfo {volume} {306}},\ \bibinfo {pages} {1910}
		(\bibinfo {year} {2004})}\BibitemShut {NoStop}%
	\bibitem [{\citenamefont {Wunderlich}\ \emph {et~al.}(2005)\citenamefont
		{Wunderlich}, \citenamefont {Kaestner}, \citenamefont {Sinova},\ and\
		\citenamefont {Jungwirth}}]{Wunderlich_SHE}%
	\BibitemOpen
	\bibfield  {author} {\bibinfo {author} {\bibfnamefont {J.}~\bibnamefont
			{Wunderlich}}, \bibinfo {author} {\bibfnamefont {B.}~\bibnamefont
			{Kaestner}}, \bibinfo {author} {\bibfnamefont {J.}~\bibnamefont {Sinova}}, \
		and\ \bibinfo {author} {\bibfnamefont {T.}~\bibnamefont {Jungwirth}},\ }\href
	{\doibase 10.1103/PhysRevLett.94.047204} {\bibfield  {journal} {\bibinfo
			{journal} {Phys. Rev. Lett.}\ }\textbf {\bibinfo {volume} {94}},\ \bibinfo
		{pages} {047204} (\bibinfo {year} {2005})}\BibitemShut {NoStop}%
	\bibitem [{\citenamefont {Chen}\ \emph {et~al.}(2016)\citenamefont {Chen},
		\citenamefont {Dumas}, \citenamefont {Eklund}, \citenamefont {Muduli},
		\citenamefont {Houshang}, \citenamefont {Awad}, \citenamefont {Dürrenfeld},
		\citenamefont {Malm}, \citenamefont {Rusu},\ and\ \citenamefont
		{Akerman}}]{Dumas_STO}%
	\BibitemOpen
	\bibfield  {author} {\bibinfo {author} {\bibfnamefont {T.}~\bibnamefont
			{Chen}}, \bibinfo {author} {\bibfnamefont {R.~K.}\ \bibnamefont {Dumas}},
		\bibinfo {author} {\bibfnamefont {A.}~\bibnamefont {Eklund}}, \bibinfo
		{author} {\bibfnamefont {P.~K.}\ \bibnamefont {Muduli}}, \bibinfo {author}
		{\bibfnamefont {A.}~\bibnamefont {Houshang}}, \bibinfo {author}
		{\bibfnamefont {A.~A.}\ \bibnamefont {Awad}}, \bibinfo {author}
		{\bibfnamefont {P.}~\bibnamefont {Dürrenfeld}}, \bibinfo {author}
		{\bibfnamefont {B.~G.}\ \bibnamefont {Malm}}, \bibinfo {author}
		{\bibfnamefont {A.}~\bibnamefont {Rusu}}, \ and\ \bibinfo {author}
		{\bibfnamefont {J.}~\bibnamefont {Akerman}},\ }\href {\doibase
		10.1109/JPROC.2016.2554518} {\bibfield  {journal} {\bibinfo  {journal}
			{Proceedings of the IEEE}\ }\textbf {\bibinfo {volume} {104}},\ \bibinfo
		{pages} {1919} (\bibinfo {year} {2016})}\BibitemShut {NoStop}%
	\bibitem [{\citenamefont {Li}\ and\ \citenamefont {Zhang}(2004)}]{ZhangLi_prb}%
	\BibitemOpen
	\bibfield  {author} {\bibinfo {author} {\bibfnamefont {Z.}~\bibnamefont
			{Li}}\ and\ \bibinfo {author} {\bibfnamefont {S.}~\bibnamefont {Zhang}},\
	}\href {\doibase 10.1103/PhysRevB.70.024417} {\bibfield  {journal} {\bibinfo
			{journal} {Phys. Rev. B}\ }\textbf {\bibinfo {volume} {70}},\ \bibinfo
		{pages} {024417} (\bibinfo {year} {2004})}\BibitemShut {NoStop}%
	\bibitem [{\citenamefont {Zhang}\ and\ \citenamefont {Li}(2004)}]{ZhangLi_prl}%
	\BibitemOpen
	\bibfield  {author} {\bibinfo {author} {\bibfnamefont {S.}~\bibnamefont
			{Zhang}}\ and\ \bibinfo {author} {\bibfnamefont {Z.}~\bibnamefont {Li}},\
	}\href {\doibase 10.1103/PhysRevLett.93.127204} {\bibfield  {journal}
		{\bibinfo  {journal} {Phys. Rev. Lett.}\ }\textbf {\bibinfo {volume} {93}},\
		\bibinfo {pages} {127204} (\bibinfo {year} {2004})}\BibitemShut {NoStop}%
	\bibitem [{\citenamefont {Brataas}\ \emph {et~al.}(2012)\citenamefont
		{Brataas}, \citenamefont {Kent},\ and\ \citenamefont {Ohno}}]{Brataas2012}%
	\BibitemOpen
	\bibfield  {author} {\bibinfo {author} {\bibfnamefont {A.}~\bibnamefont
			{Brataas}}, \bibinfo {author} {\bibfnamefont {A.~D.}\ \bibnamefont {Kent}}, \
		and\ \bibinfo {author} {\bibfnamefont {H.}~\bibnamefont {Ohno}},\ }\href
	{\doibase 10.1038/nmat3311} {\bibfield  {journal} {\bibinfo  {journal}
			{Nature Materials}\ }\textbf {\bibinfo {volume} {11}},\ \bibinfo {pages}
		{372} (\bibinfo {year} {2012})}\BibitemShut {NoStop}%
	\bibitem [{\citenamefont {Consolo}\ \emph {et~al.}(2007)\citenamefont
		{Consolo}, \citenamefont {Azzerboni}, \citenamefont {Gerhart}, \citenamefont
		{Melkov}, \citenamefont {Tiberkevich},\ and\ \citenamefont
		{Slavin}}]{Consolo_prb2007}%
	\BibitemOpen
	\bibfield  {author} {\bibinfo {author} {\bibfnamefont {G.}~\bibnamefont
			{Consolo}}, \bibinfo {author} {\bibfnamefont {B.}~\bibnamefont {Azzerboni}},
		\bibinfo {author} {\bibfnamefont {G.}~\bibnamefont {Gerhart}}, \bibinfo
		{author} {\bibfnamefont {G.~A.}\ \bibnamefont {Melkov}}, \bibinfo {author}
		{\bibfnamefont {V.}~\bibnamefont {Tiberkevich}}, \ and\ \bibinfo {author}
		{\bibfnamefont {A.~N.}\ \bibnamefont {Slavin}},\ }\href {\doibase
		10.1103/PhysRevB.76.144410} {\bibfield  {journal} {\bibinfo  {journal} {Phys.
				Rev. B}\ }\textbf {\bibinfo {volume} {76}},\ \bibinfo {pages} {144410}
		(\bibinfo {year} {2007})}\BibitemShut {NoStop}%
	\bibitem [{\citenamefont {{Slavin}}\ and\ \citenamefont
		{{Tiberkevich}}(2009)}]{Slavin_ieee}%
	\BibitemOpen
	\bibfield  {author} {\bibinfo {author} {\bibfnamefont {A.}~\bibnamefont
			{{Slavin}}}\ and\ \bibinfo {author} {\bibfnamefont {V.}~\bibnamefont
			{{Tiberkevich}}},\ }\href@noop {} {\bibfield  {journal} {\bibinfo  {journal}
			{IEEE Transactions on Magnetics}\ }\textbf {\bibinfo {volume} {45}},\
		\bibinfo {pages} {1875} (\bibinfo {year} {2009})}\BibitemShut {NoStop}%
	\bibitem [{\citenamefont {Hoefer}\ \emph {et~al.}(2010)\citenamefont {Hoefer},
		\citenamefont {Silva},\ and\ \citenamefont {Keller}}]{Hoefer2010}%
	\BibitemOpen
	\bibfield  {author} {\bibinfo {author} {\bibfnamefont {M.~A.}\ \bibnamefont
			{Hoefer}}, \bibinfo {author} {\bibfnamefont {T.~J.}\ \bibnamefont {Silva}}, \
		and\ \bibinfo {author} {\bibfnamefont {M.~W.}\ \bibnamefont {Keller}},\
	}\href {\doibase 10.1103/PhysRevB.82.054432} {\bibfield  {journal} {\bibinfo
			{journal} {Phys. Rev. B}\ }\textbf {\bibinfo {volume} {82}},\ \bibinfo
		{pages} {054432} (\bibinfo {year} {2010})}\BibitemShut {NoStop}%
	\bibitem [{\citenamefont {Chung}\ \emph {et~al.}(2018)\citenamefont {Chung},
		\citenamefont {Le}, \citenamefont {Ahlberg}, \citenamefont {Awad},
		\citenamefont {Weigand}, \citenamefont {Bykova}, \citenamefont {Khymyn},
		\citenamefont {Dvornik}, \citenamefont {Mazraati}, \citenamefont {Houshang},
		\citenamefont {Jiang}, \citenamefont {Nguyen}, \citenamefont {Goering},
		\citenamefont {Sch\"utz}, \citenamefont {Gr\"afe},\ and\ \citenamefont
		{\AA{}kerman}}]{AkermanXMCD}%
	\BibitemOpen
	\bibfield  {author} {\bibinfo {author} {\bibfnamefont {S.}~\bibnamefont
			{Chung}}, \bibinfo {author} {\bibfnamefont {Q.~T.}\ \bibnamefont {Le}},
		\bibinfo {author} {\bibfnamefont {M.}~\bibnamefont {Ahlberg}}, \bibinfo
		{author} {\bibfnamefont {A.~A.}\ \bibnamefont {Awad}}, \bibinfo {author}
		{\bibfnamefont {M.}~\bibnamefont {Weigand}}, \bibinfo {author} {\bibfnamefont
			{I.}~\bibnamefont {Bykova}}, \bibinfo {author} {\bibfnamefont
			{R.}~\bibnamefont {Khymyn}}, \bibinfo {author} {\bibfnamefont
			{M.}~\bibnamefont {Dvornik}}, \bibinfo {author} {\bibfnamefont
			{H.}~\bibnamefont {Mazraati}}, \bibinfo {author} {\bibfnamefont
			{A.}~\bibnamefont {Houshang}}, \bibinfo {author} {\bibfnamefont
			{S.}~\bibnamefont {Jiang}}, \bibinfo {author} {\bibfnamefont {T.~N.~A.}\
			\bibnamefont {Nguyen}}, \bibinfo {author} {\bibfnamefont {E.}~\bibnamefont
			{Goering}}, \bibinfo {author} {\bibfnamefont {G.}~\bibnamefont {Sch\"utz}},
		\bibinfo {author} {\bibfnamefont {J.}~\bibnamefont {Gr\"afe}}, \ and\
		\bibinfo {author} {\bibfnamefont {J.}~\bibnamefont {\AA{}kerman}},\ }\href
	{\doibase 10.1103/PhysRevLett.120.217204} {\bibfield  {journal} {\bibinfo
			{journal} {Phys. Rev. Lett.}\ }\textbf {\bibinfo {volume} {120}},\ \bibinfo
		{pages} {217204} (\bibinfo {year} {2018})}\BibitemShut {NoStop}%
	\bibitem [{\citenamefont {Backes}\ \emph {et~al.}(2015)\citenamefont {Backes},
		\citenamefont {Maci\`a}, \citenamefont {Bonetti}, \citenamefont {Kukreja},
		\citenamefont {Ohldag},\ and\ \citenamefont {Kent}}]{Backes2015}%
	\BibitemOpen
	\bibfield  {author} {\bibinfo {author} {\bibfnamefont {D.}~\bibnamefont
			{Backes}}, \bibinfo {author} {\bibfnamefont {F.}~\bibnamefont {Maci\`a}},
		\bibinfo {author} {\bibfnamefont {S.}~\bibnamefont {Bonetti}}, \bibinfo
		{author} {\bibfnamefont {R.}~\bibnamefont {Kukreja}}, \bibinfo {author}
		{\bibfnamefont {H.}~\bibnamefont {Ohldag}}, \ and\ \bibinfo {author}
		{\bibfnamefont {A.~D.}\ \bibnamefont {Kent}},\ }\href {\doibase
		10.1103/PhysRevLett.115.127205} {\bibfield  {journal} {\bibinfo  {journal}
			{Phys. Rev. Lett.}\ }\textbf {\bibinfo {volume} {115}},\ \bibinfo {pages}
		{127205} (\bibinfo {year} {2015})}\BibitemShut {NoStop}%
	\bibitem [{\citenamefont {Zhou}\ \emph {et~al.}(2015)\citenamefont {Zhou},
		\citenamefont {Iacocca}, \citenamefont {Awad}, \citenamefont {Dumas},
		\citenamefont {Zhang}, \citenamefont {Braun},\ and\ \citenamefont
		{{\AA}kerman}}]{Zhou_natComm_2015}%
	\BibitemOpen
	\bibfield  {author} {\bibinfo {author} {\bibfnamefont {Y.}~\bibnamefont
			{Zhou}}, \bibinfo {author} {\bibfnamefont {E.}~\bibnamefont {Iacocca}},
		\bibinfo {author} {\bibfnamefont {A.~A.}\ \bibnamefont {Awad}}, \bibinfo
		{author} {\bibfnamefont {R.~K.}\ \bibnamefont {Dumas}}, \bibinfo {author}
		{\bibfnamefont {F.~C.}\ \bibnamefont {Zhang}}, \bibinfo {author}
		{\bibfnamefont {H.~B.}\ \bibnamefont {Braun}}, \ and\ \bibinfo {author}
		{\bibfnamefont {J.}~\bibnamefont {{\AA}kerman}},\ }\href {\doibase
		10.1038/ncomms9193} {\bibfield  {journal} {\bibinfo  {journal} {Nat.
				Commun.}\ }\textbf {\bibinfo {volume} {6}},\ \bibinfo {pages} {8193}
		(\bibinfo {year} {2015})}\BibitemShut {NoStop}%
	\bibitem [{\citenamefont {Liu}\ \emph {et~al.}(2015)\citenamefont {Liu},
		\citenamefont {Lim},\ and\ \citenamefont {Urazhdin}}]{Sergei2015}%
	\BibitemOpen
	\bibfield  {author} {\bibinfo {author} {\bibfnamefont {R.~H.}\ \bibnamefont
			{Liu}}, \bibinfo {author} {\bibfnamefont {W.~L.}\ \bibnamefont {Lim}}, \ and\
		\bibinfo {author} {\bibfnamefont {S.}~\bibnamefont {Urazhdin}},\ }\href
	{\doibase 10.1103/PhysRevLett.114.137201} {\bibfield  {journal} {\bibinfo
			{journal} {Phys. Rev. Lett.}\ }\textbf {\bibinfo {volume} {114}},\ \bibinfo
		{pages} {137201} (\bibinfo {year} {2015})}\BibitemShut {NoStop}%
	\bibitem [{\citenamefont {Statuto}\ \emph {et~al.}(2018)\citenamefont
		{Statuto}, \citenamefont {Hern{\`{a}}ndez}, \citenamefont {Kent},\ and\
		\citenamefont {Maci{\`{a}}}}]{Statuto_2018}%
	\BibitemOpen
	\bibfield  {author} {\bibinfo {author} {\bibfnamefont {N.}~\bibnamefont
			{Statuto}}, \bibinfo {author} {\bibfnamefont {J.~M.}\ \bibnamefont
			{Hern{\`{a}}ndez}}, \bibinfo {author} {\bibfnamefont {A.~D.}\ \bibnamefont
			{Kent}}, \ and\ \bibinfo {author} {\bibfnamefont {F.}~\bibnamefont
			{Maci{\`{a}}}},\ }\href {\doibase 10.1088/1361-6528/aac411} {\bibfield
		{journal} {\bibinfo  {journal} {Nanotechnology}\ }\textbf {\bibinfo {volume}
			{29}},\ \bibinfo {pages} {325302} (\bibinfo {year} {2018})}\BibitemShut
	{NoStop}%
	\bibitem [{\citenamefont {Vansteenkiste}\ \emph {et~al.}(2014)\citenamefont
		{Vansteenkiste}, \citenamefont {Leliaert}, \citenamefont {Dvornik},
		\citenamefont {Helsen}, \citenamefont {Garcia-Sanchez},\ and\ \citenamefont
		{Waeyenberge}}]{Vansteenkiste-mumax}%
	\BibitemOpen
	\bibfield  {author} {\bibinfo {author} {\bibfnamefont {A.}~\bibnamefont
			{Vansteenkiste}}, \bibinfo {author} {\bibfnamefont {J.}~\bibnamefont
			{Leliaert}}, \bibinfo {author} {\bibfnamefont {M.}~\bibnamefont {Dvornik}},
		\bibinfo {author} {\bibfnamefont {M.}~\bibnamefont {Helsen}}, \bibinfo
		{author} {\bibfnamefont {F.}~\bibnamefont {Garcia-Sanchez}}, \ and\ \bibinfo
		{author} {\bibfnamefont {B.~V.}\ \bibnamefont {Waeyenberge}},\ }\href
	{\doibase 10.1063/1.4899186} {\bibfield  {journal} {\bibinfo  {journal} {AIP
				Advances}\ }\textbf {\bibinfo {volume} {4}},\ \bibinfo {pages} {107133}
		(\bibinfo {year} {2014})}\BibitemShut {NoStop}%
	\bibitem [{\citenamefont {Mancoff}\ \emph {et~al.}(2006)\citenamefont
		{Mancoff}, \citenamefont {Rizzo}, \citenamefont {Engel},\ and\ \citenamefont
		{Tehrani}}]{Mancoff2006}%
	\BibitemOpen
	\bibfield  {author} {\bibinfo {author} {\bibfnamefont {F.~B.}\ \bibnamefont
			{Mancoff}}, \bibinfo {author} {\bibfnamefont {N.~D.}\ \bibnamefont {Rizzo}},
		\bibinfo {author} {\bibfnamefont {B.~N.}\ \bibnamefont {Engel}}, \ and\
		\bibinfo {author} {\bibfnamefont {S.}~\bibnamefont {Tehrani}},\ }\href
	{https://doi.org/10.1063/1.2185620} {\bibfield  {journal} {\bibinfo
			{journal} {Applied Physics Letters}\ }\textbf {\bibinfo {volume} {88}},\
		\bibinfo {pages} {112507} (\bibinfo {year} {2006})}\BibitemShut {NoStop}%
	\bibitem [{\citenamefont {Bonetti}\ \emph {et~al.}(2010)\citenamefont
		{Bonetti}, \citenamefont {Tiberkevich}, \citenamefont {Consolo},
		\citenamefont {Finocchio}, \citenamefont {Muduli}, \citenamefont {Mancoff},
		\citenamefont {Slavin},\ and\ \citenamefont {{\AA}kerman}}]{Bonetti2010}%
	\BibitemOpen
	\bibfield  {author} {\bibinfo {author} {\bibfnamefont {S.}~\bibnamefont
			{Bonetti}}, \bibinfo {author} {\bibfnamefont {V.}~\bibnamefont
			{Tiberkevich}}, \bibinfo {author} {\bibfnamefont {G.}~\bibnamefont
			{Consolo}}, \bibinfo {author} {\bibfnamefont {G.}~\bibnamefont {Finocchio}},
		\bibinfo {author} {\bibfnamefont {P.}~\bibnamefont {Muduli}}, \bibinfo
		{author} {\bibfnamefont {F.}~\bibnamefont {Mancoff}}, \bibinfo {author}
		{\bibfnamefont {A.}~\bibnamefont {Slavin}}, \ and\ \bibinfo {author}
		{\bibfnamefont {J.}~\bibnamefont {{\AA}kerman}},\ }\href {\doibase
		10.1103/PhysRevLett.105.217204} {\bibfield  {journal} {\bibinfo  {journal}
			{Phys. Rev. Lett.}\ }\textbf {\bibinfo {volume} {105}},\ \bibinfo {pages}
		{217204} (\bibinfo {year} {2010})}\BibitemShut {NoStop}%
	\bibitem [{\citenamefont {Hang}\ \emph {et~al.}(2018)\citenamefont {Hang},
		\citenamefont {Hahn}, \citenamefont {Statuto}, \citenamefont {Maci\`a},\ and\
		\citenamefont {Kent}}]{Jinting_2017}%
	\BibitemOpen
	\bibfield  {author} {\bibinfo {author} {\bibfnamefont {J.}~\bibnamefont
			{Hang}}, \bibinfo {author} {\bibfnamefont {C.}~\bibnamefont {Hahn}}, \bibinfo
		{author} {\bibfnamefont {N.}~\bibnamefont {Statuto}}, \bibinfo {author}
		{\bibfnamefont {F.}~\bibnamefont {Maci\`a}}, \ and\ \bibinfo {author}
		{\bibfnamefont {A.~D.}\ \bibnamefont {Kent}},\ }\href {\doibase
		10.1038/s41598-018-25134-z} {\bibfield  {journal} {\bibinfo  {journal}
			{Scientific Reports}\ ,\ \bibinfo {pages} {6847}} (\bibinfo {year}
		{2018})}\BibitemShut {NoStop}%
	\bibitem [{\citenamefont {Giordano}\ \emph {et~al.}(2014)\citenamefont
		{Giordano}, \citenamefont {Carpentieri}, \citenamefont {Laudani},
		\citenamefont {Gubbiotti}, \citenamefont {Azzerboni},\ and\ \citenamefont
		{Finocchio}}]{simulation_shno}%
	\BibitemOpen
	\bibfield  {author} {\bibinfo {author} {\bibfnamefont {A.}~\bibnamefont
			{Giordano}}, \bibinfo {author} {\bibfnamefont {M.}~\bibnamefont
			{Carpentieri}}, \bibinfo {author} {\bibfnamefont {A.}~\bibnamefont
			{Laudani}}, \bibinfo {author} {\bibfnamefont {G.}~\bibnamefont {Gubbiotti}},
		\bibinfo {author} {\bibfnamefont {B.}~\bibnamefont {Azzerboni}}, \ and\
		\bibinfo {author} {\bibfnamefont {G.}~\bibnamefont {Finocchio}},\ }\href
	{https://doi.org/10.1063/1.4892168} {\bibfield  {journal} {\bibinfo
			{journal} {Applied Physics Letters}\ }\textbf {\bibinfo {volume} {105}},\
		\bibinfo {pages} {042412} (\bibinfo {year} {2014})}\BibitemShut {NoStop}%
\end{thebibliography}
%

\end{document}